\documentclass{PoS}
\pdfoutput=1
\usepackage{rotating,booktabs,amsmath,amssymb, multicol,url}
\usepackage[numbers,sort&compress]{natbib}
\usepackage{natbibspacing}
\newcommand{\comment}[1]{}

\providecommand{\abs}[1]{\lvert#1\rvert}    %        |.|
\title{$B$-meson decay constants with domain-wall light quarks and nonperturbatively tuned relativistic $b$-quarks}

\ShortTitle{$B$-meson decay constants}

\author{\speaker{Oliver Witzel}\\%
        Center for Computational Science, Boston University,\newline 3 Cummington Mall, Boston, MA 02215, USA\\
        E-mail: \email{owitzel@bu.edu}}

\abstract{We report on our progress to obtain the decay constants $f_B$ and $f_{B_s}$ from lattice-QCD simulations on the RBC-UKQCD Collaborations 2+1 flavor domain-wall Iwasaki lattices.  Using domain-wall light quarks and relativistic $b$-quarks we analyze data with several partially quenched light-quark masses at two lattice spacings of $a \approx 0.11$ fm and $a \approx 0.08$ fm. % We perform combined chiral and continuum extrapolations using chiral perturbation theory.
}

\FullConference{31st International Symposium on Lattice Field Theory - LATTICE 2013\\
                July 29 - August 3, 2013\\
                Mainz, Germany}

\begin{document}

\section{Motivation}
\begin{figure}[tb]
\centering
\includegraphics[scale=1.0]{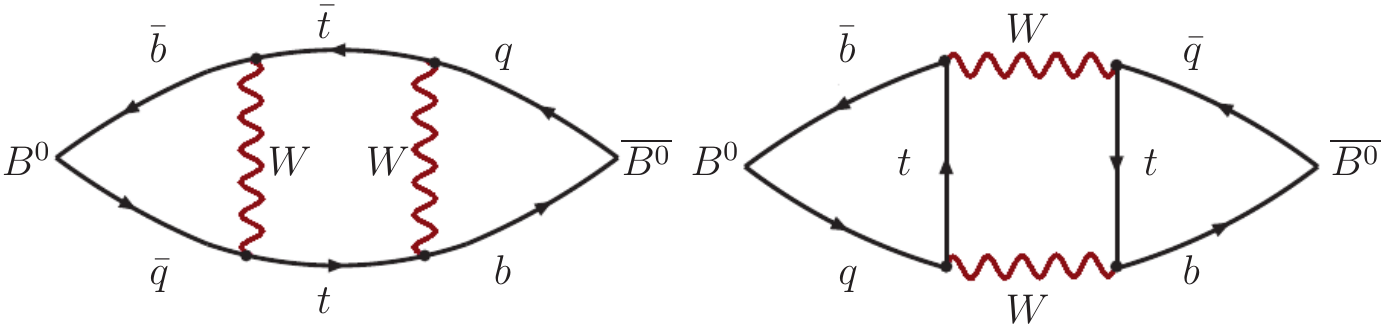}
\caption{Box-diagrams with top-quarks in the loop are the dominant contributions to neutral $B$-meson mixing. $q$ denotes either a $d$- or $s$-quark.}
\label{Fig:Mixing}
\end{figure}

$B$-physics plays a central role in the global efforts to constrain the CKM unitarity triangle. The ratio of neutral $B$-meson mixing, e.g., is used in the unitarity triangle fits  \cite{Charles:2004jd,Bona:2005vz,Laiho:2009eu}. Neutral $B$-mesons mix with their anti-particle under the exchange of two $W$-bosons as depicted by the box-diagrams in Fig.~\ref{Fig:Mixing}. There $q$ denotes a light $d$- or $s$-quark building either a $B$- or a $B_s$-meson, respectively. In the experiments, e.g., BaBar, Belle, CDF or LHCb, $B_q$-mixing is measured in terms of the oscillation frequencies (mass differences) $\Delta M_q$ and in the Standard Model (SM) this process is parameterized by \cite{Buras:1990fn}
\begin{align}
\Delta M_q = \frac{G_F^2m^2_W}{6\pi^2} \eta_B S_0 M_{B_q}{f_{B_q}^2B_{B_q}} \abs{V_{tq}^*V_{tb}}^2,
\end{align}
where the QCD coefficient $\eta_b$ \cite{Buras:1990fn} and the Inami-Lim function $S_0$ \cite{Inami:1980fz} are computed perturbatively and a nonperturbative computation is needed for the leptonic $B_q$-meson decay constant $f_{B_q}$ and the bag parameter $B_{B_q}$ in order to extract the CKM matrix elements $V^*_{tq}V_{tb}$. Experimentally $\Delta M_q$ is measured to subpercent accuracy \cite{Beringer:2012zz}, whereas the nonpeturbative (lattice) inputs contribute the dominant uncertainty (order few percent). Taking the ratio of neutral $B$-meson mixing 
\begin{align}
\frac{\Delta M_s}{\Delta M_d} = \frac{M_{B_s}}{M_{B_d}}\,{\xi^2} \, \frac{\abs{V_{ts}}^2}{\abs{V_{td}}^2},
\end{align}
the nonperturbative contribution is contained in the $SU(3)$ breaking ratio 
\begin{align}
\xi &= \frac{f_{B_s}\sqrt{B_{B_s}}}{f_{B_d}\sqrt{B_{B_d}}},
\end{align}
for which statistical and systematic uncertainties largely cancel \cite{Bernard:1998dg}. Unfortunately $\xi$ still contributes the largest uncertainty.

We therefore designed this project to compute neutral $B$-meson mixing matrix elements as well as the leptonic decay constants $f_B$ and $f_{B_s}$. The decay constants are important to further constrain new physics by allowing an alternative determination of $V_{ub}$ using the measurement of $B\to\tau \nu$ \cite{Lunghi:2009ke,Bona:2009cj,Lenz:2010gu} or by allowing, e.g., to obtain predictions on rare decays like $B_s \to \mu_+ \mu_-$  \cite{Buras:2013uqa} which promise to be in particular sensitive to new physics. %The SM prediction for the $BR(B\to \tau \nu)$ needs $f_B$ as input. For $B\to\tau\nu$ flavor-changing neutral currents (FCNC) are suppressed by the Glashow-Iliopoulos-Maiani (GIM)-mechanism and charged current decays are helicity suppressed. This leads to a potential sensitivity to tree-level effects of new scalar particles like a charged Higgs boson as it is introduced in multi-Higgs extensions of the SM (e.g.~type-II two Higgs doublet model or the MSSM) \cite{Bona:2009cj}. Likewise the SM prediction for the $BR(B_s \to \mu_+\mu_-)$ needs $f_{B_s}$ as input \cite{Buras:2012ub,Buras:2013uqa}. Recently LHCb measured this decay with $3.5\sigma$ significance \cite{Aaij:2012nna} and combined with CMS results a significance of greater $5\sigma$ is found \cite{Hansmann-Menzemer:2013eps}. The current result is in agreement with the SM. 

Computing $B$-physics quantities on the lattice faces the additional challenge to accommodate an additional scale given by the large $b$-quark mass. In our project we compute $B$-physics quantities using the RBC-UKQCD 2+1 flavor domain-wall Iwasaki gauge field configurations. We simulate the $b$-quarks with the relativistic heavy quark (RHQ) action and tune the action's parameters nonperturbatively, while domain-wall fermions simulate the light $u,\,d,\,s$-quarks. Thus our project is an independent cross-check to published results by other groups based on 2-flavor \cite{Dimopoulos:2011gx}, %Blossier:2011dk,Bernardoni:2012ti,Carrasco:2012de 
2+1-flavor \cite{Gamiz:2009ku,McNeile:2011ng,Na:2012kp,Bazavov:2011aa,Bazavov:2012zs} or 2+1+1-flavor \cite{Dowdall:2013tga} gauge-field configurations. In these proceedings we focus on the computation of the $B$-meson decay constants $f_B$ and $f_{B_s}$.

\section{Computational setup}
This computation uses the dynamical 2+1 flavor domain-wall Iwasaki gaugefield configurations generated by the RBC-UKQCD collaboration \cite{Allton:2008pn,Aoki:2010dy} listed in Tab.~\ref{tab:lattices}. We use two coarser, $24^3$ ensembles with $a \approx 0.11 $fm ($a^{-1} = 1.729$ GeV) and three finer, $32^3$ ensembles with $a \approx 0.086$ fm ($a^{-1} = 2.281$ GeV). On the coarser ensembles we place one source per configuration, whereas on the finer ensembles we place two time sources per configuration separated by half the temporal extent of the lattice. For each source we generate six domain-wall \cite{Kaplan:1992bt,Shamir:1993zy} propagators with quark masses $a m_\text{val}^{24}$ = 0.005, 0.010, 0.020, 0.030, 0.0343 and $0.040$ on the coarser $24^3$ ensembles and $a m_\text{val}^{32}$ = 0.004, 0.006, 0.008, 0.025, 0.0272 and 0.030 on the finer $32^3$ ensembles. The masses of the three heaviest domain-wall propagators bracket the physical strange quark mass. %which on the corresponding ensembles is found to be $a m_s^{24} = 0.0348(11)$ and $a m_s^{32} = 0.0273(11)$ \cite{Aoki:2010dy}. %In the following we consider the two (three) lightest $24^3$ ($32^3$) domain-wall propagators to be in the ``chiral'' regime.
\begin{table}[t]
\centering
\caption{Lattice simulation parameters used in our $B$-physics program.  The columns list the lattice volume, approximate lattice spacing, light ($m_l$) and strange ($m_h$) sea-quark masses, unitary pion mass, and number of configurations and time sources analyzed.}
\vspace{3mm}
\label{tab:lattices}
\begin{tabular}{cccccrc} \toprule
%  &         &         &         &            & &\# time\\
$\left(L/a\right)^3 \times \left(T/a\right)$ \qquad & $\approx a$(fm) & ~~$am_l$ & ~~$am_h$ & \quad $M_\pi$(MeV) \quad & \# configs.&\# time sources\\[0.5mm] \midrule
$24^3 \times 64$ &  0.11 &  0.005 & 0.040 & 329 & 1636~~~~&1\\
$24^3 \times 64$ &  0.11 &  0.010 & 0.040 & 422 & 1419~~~~&1\\ \midrule
$32^3 \times 64$ &  0.086 &  0.004 & 0.030 & 289 & 628~~~~&2\\ 
$32^3 \times 64$ &  0.086 &  0.006 & 0.030 & 345 & 889~~~~&2\\
$32^3 \times 64$ &  0.086 &  0.008 & 0.030 & 394 & 544~~~~&2\\ \bottomrule
\end{tabular}\end{table}

We simulate the $b$-quarks using the the anisotropic Sheikholeslami-Wohlert (clover) action with the relativistic heavy-quark (RHQ) interpretation~\cite{Christ:2006us,Lin:2006ur}. The three parameters, $m_0a$, $c_P$, $\zeta$, are tuned nonperturbatively using the experimental inputs for the spin-averaged mass $\overline M$ and the hyperfine-splitting  $\Delta_M$  in the $B_s$-meson system and demanding that the rest mass equals the kinetic mass, i.e., $M_1/M_2=1$ \cite{Aoki:2012xaa}. The parameters are tuned by probing seven points of the $(m_0a,\,c_P,\,\zeta)$ parameter space and then we interpolate to the tuned value by matching to the experimental values. 

%\begin{figure}[tb]
%\centering
%\includegraphics[scale=0.45]{images/RHQparameters}
%\caption{The seven RHQ parameter sets are indicated by the six endpoints and the central point of the molecule-like structure in the $m_0a$, $c_P$ and $\zeta$ parameter space.}
%\label{Fig:RHQparameters}
%\end{figure}

\begin{figure}[tb]
\centering
\begin{minipage}{0.48\textwidth}
\includegraphics[scale=0.5]{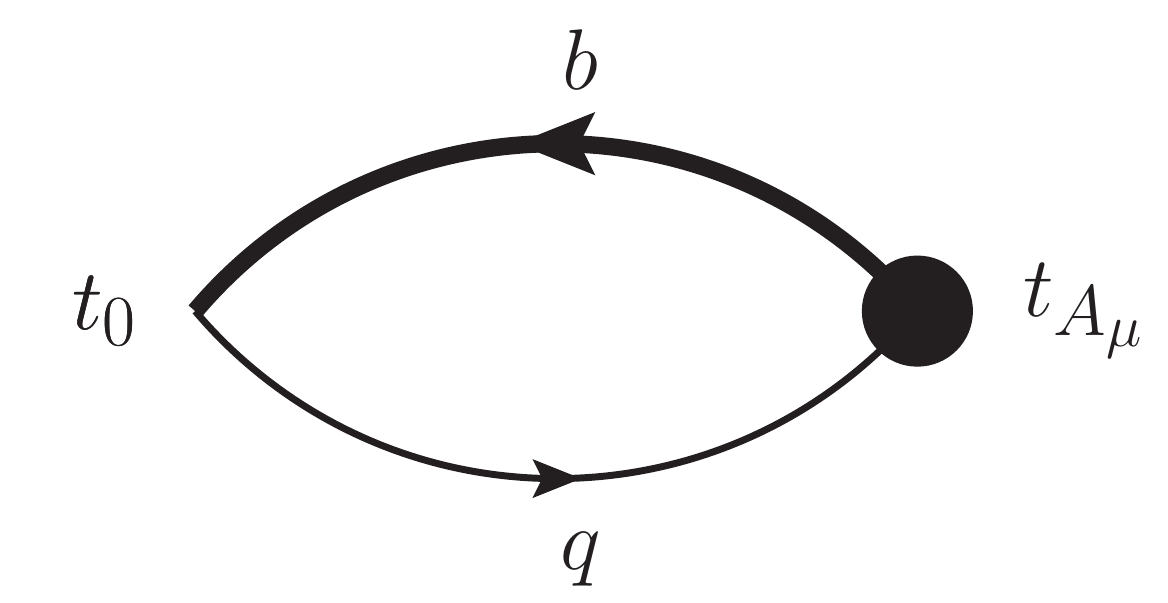}
\caption{Schematic computation of the decay amplitude $\Phi_{B_q}$ with $q$ denoting a $d$- or $s$-quark.}
\label{Fig:fB}
\end{minipage}
\hfill
\begin{minipage}{0.48\textwidth}
\includegraphics[scale=0.5]{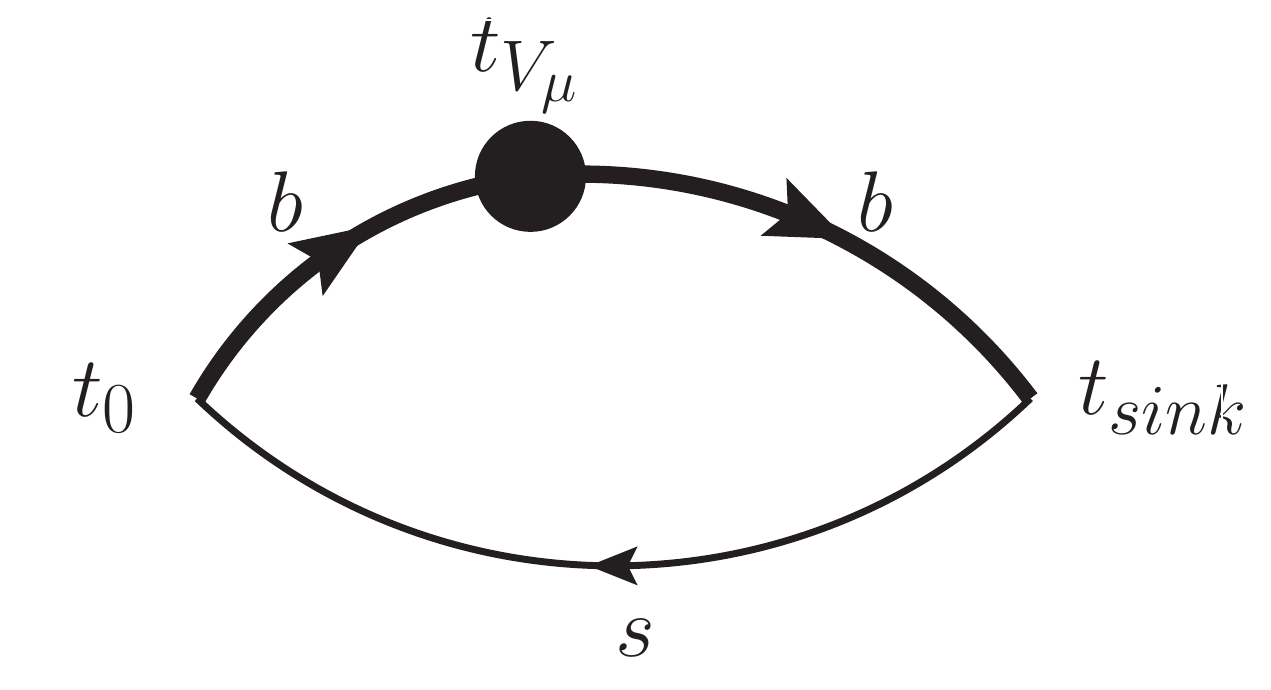}
\caption{Schematic computation of the flavor-conserving renormalization factor $Z_V^{bb}$ using a $s$-quark as spectator.}
\label{Fig:Zvbb}
\end{minipage}
\end{figure}

We use the same seven sets of RHQ parameters in our computation of the decay constants $f_B$ and $f_{B_s}$ because this allows us to cleanly propagate the statistical uncertainty of our tuning procedure to the final results. The decay constants are measured on the lattice by computing the decay amplitude $\Phi_B$ which is proportional to the vacuum-to-meson matrix element of the heavy-light axial  vector current ${\cal A}_\mu = \bar{b} \gamma_5 \gamma_\mu q$ and depicted in Fig.~\ref{Fig:fB}
\begin{align}
\langle 0| {\cal A}_\mu | B_q(p)\rangle / \sqrt{M_{B_q}} = i p^\mu \Phi_{B_q}^{(0)} / M_{B_q}.
\end{align}
The mass of the $B_q$-meson is $M_{B_q}$ and $p^\mu$ denotes its four momentum. We reduce lattice discretization errors by $O(a)$-improving the axial vector current,
%\begin{align}
$\Phi_{B_q}^\text{imp} = \Phi_{B_q}^{(0)} + c_1 \Phi_{B_q}^{(1)}$,
%\end{align}
and compute the coefficient $c_1$ at 1-loop with mean-field improved lattice perturbation theory \cite{Lehner:2012bt}. 

Finally we obtain the decay constant $f_{B_q}$ from $\Phi_{B_q}^\text{imp}$ by multiplying the renormalization factor $Z_\Phi$, the lattice spacing and the mass of the $B_q$-meson
\begin{align}
f_{B_q} = Z_\Phi \Phi_{B_q}^\text{imp} a^{-3/2} / \sqrt{M_{B_q}}.
\end{align}
For the computation of the renormalization factor $Z_\Phi$ we follow the mostly nonperturbative method described in \cite{ElKhadra:2001rv} and compute $Z_\Phi$  as product of the two nonperturbatively computed, flavor-conserving factors $Z_V^{ll}$ and $Z_V^{bb}$ and a perturbatively computed factor $\varrho_{bl}$  which is expected to be close to one and to have a more convergent series expansion in $\alpha_s$
\begin{align}
Z_\Phi = \varrho_{bl} \sqrt{Z_V^{bb} Z_V^{ll}}.
\end{align} 
The perturbative factor $\rho_{bl}$ is computed at 1-loop with mean-field improved lattice perturbation theory \cite{Lepage:1992xa} and the RBC-UKQCD collaboration already measured $Z_V^{ll}$ \cite{Aoki:2010dy}. The factor  $Z_V^{bb}$ is determined as part of this project \cite{Kawanai:2012id}.

\section{Preliminary results}
We determine $Z_v^{bb}$ by measuring the 3-point function describing a $B$-meson going to a $B$-meson with the insertion of a vector current between both $b$-quarks (see Fig.~\ref{Fig:Zvbb})
\begin{align}
Z_V^{bb} \times \langle B | V^{bb,0} | B\rangle = 2 m_{B}.
\end{align}
Since $Z_V^{bb}$ does not explicitly depend on the spectator quark, it is advantageous to use a $s$-quark as spectator because it has smaller statistical uncertainties compared to a lighter quark. For this computation we simulate the $b$-quarks using a single set of tuned RHQ parameters \cite{Aoki:2012xaa}.

\begin{figure}[t]
\begin{minipage}{0.45\textwidth}
\includegraphics[scale=0.37]{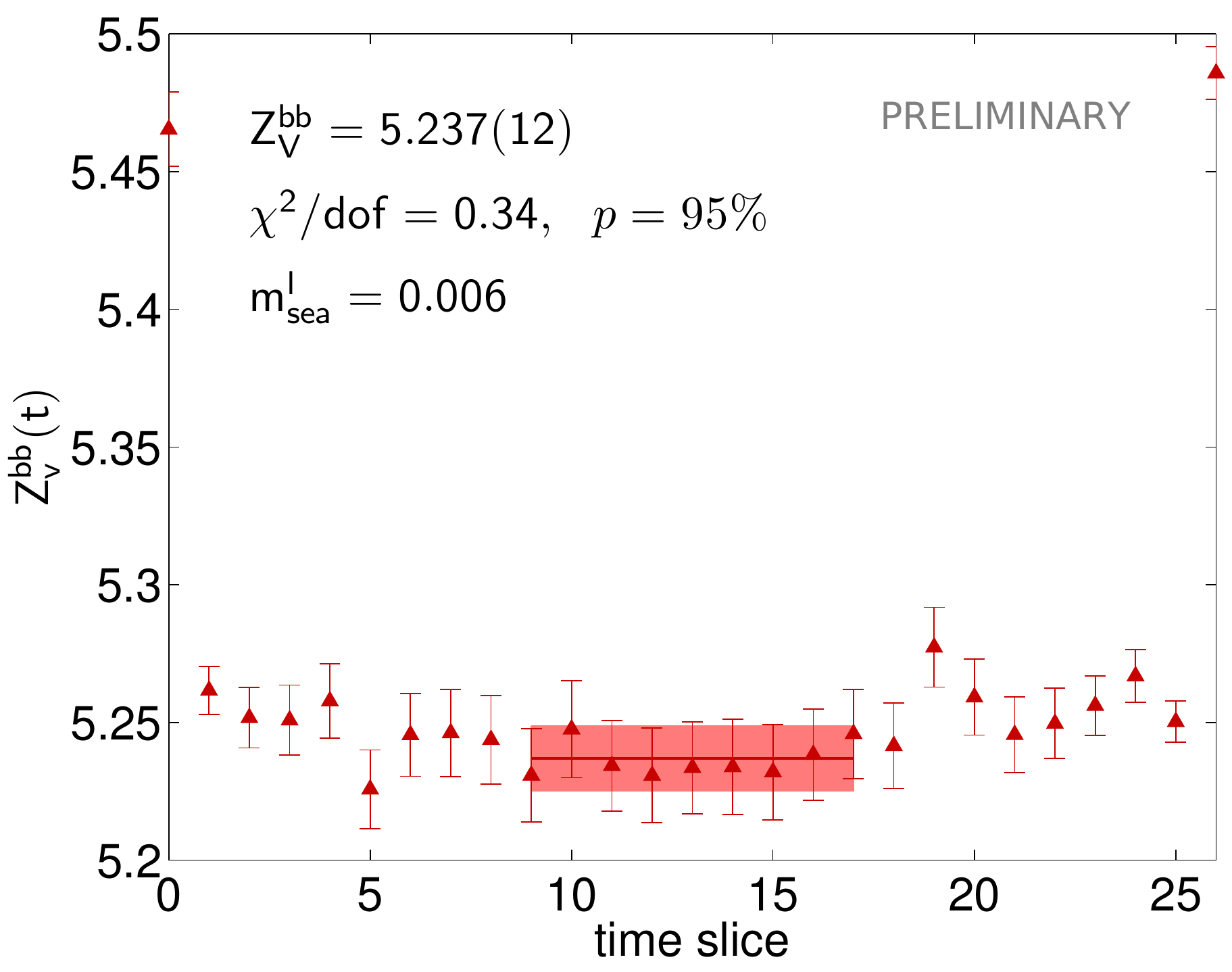}
\caption{Example plot for the determination of the flavor-conserving renormalization factor $Z_V^{bb}$ on the finer, $32^3$ ensemble with $m_\text{sea}^l = 0.006$.}
\label{Fig:Zvbb006}
\end{minipage}
\end{figure}

We extract $Z_V^{bb}$ from a fit to the plateau of the above defined 3-pt function normalized by the corresponding $B_s$-meson 2-pt function for each of our five ensembles. Fig.~\ref{Fig:Zvbb006} shows example data for $Z_V^{bb}$ on the finer, $32^3$ ensemble with light sea-quark mass $a_{32} m_\text{sea}^l = 0.006$. The data form a long plateau and the fit interval is chosen such that excited state contamination present in the 2pt-data has decayed and is not affecting our signal. Plots for the other ensembles look similar. We list the values for $Z_V^{bb}$ for all our ensembles in Tab.~\ref{Tab:Zvbb}. As expected we do not observe a dependence on the sea-quark mass. Furthermore we use the results to test the reliability of lattice perturbation theory used for different parts of this project, e.g., the factor $\varrho_{bl}$. We show the results for $Z_V^{bb}$ obtained at 1-loop mean-field improved lattice perturbation theory \cite{Lehner:2012bt} and compare them to the averages of our nonperturbative determinations. We observe a better-than-expected agreement. \\

\begin{figure}[t]
%\centering
\begin{minipage}{0.48\textwidth}
\includegraphics[scale=0.38]{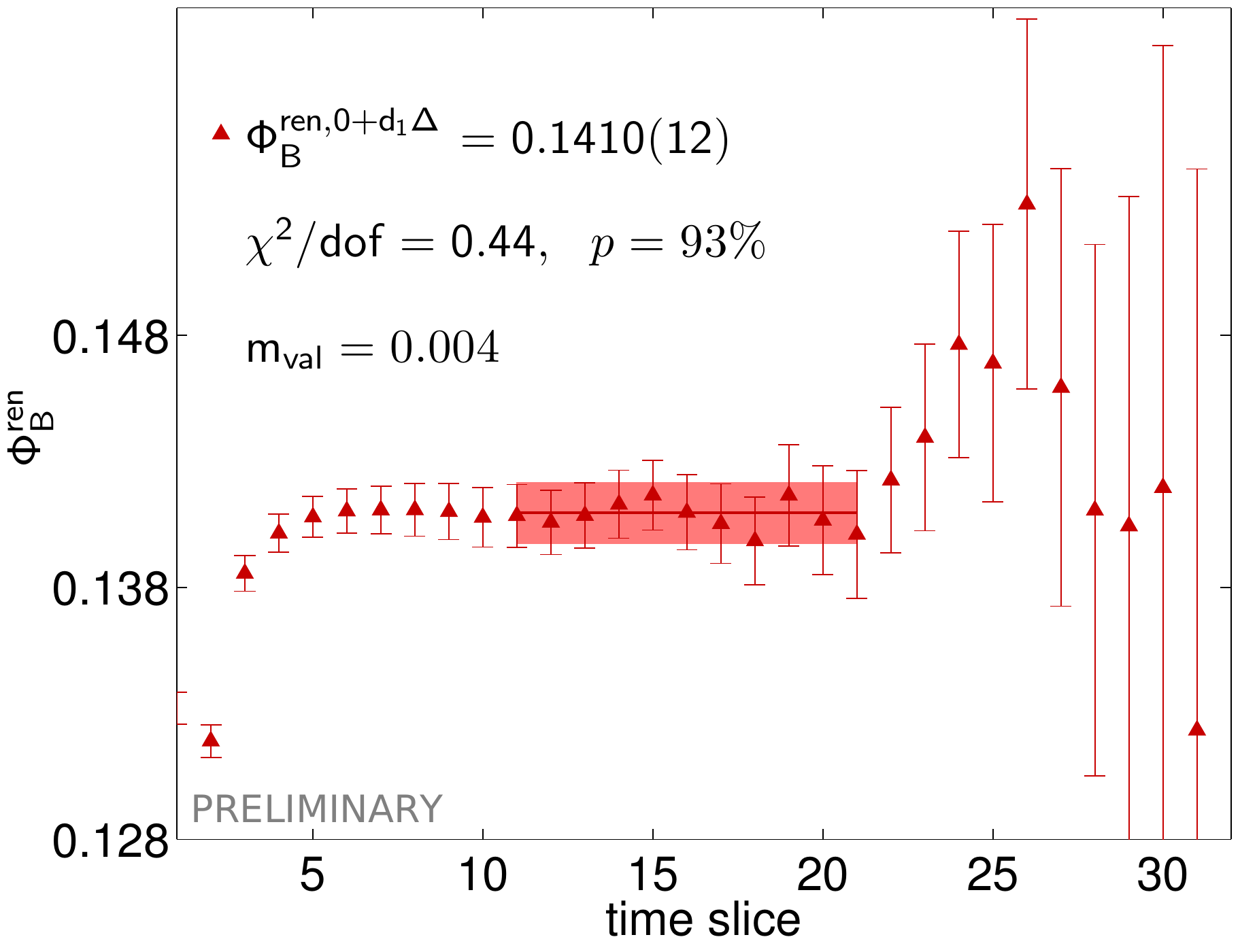}
\caption{Example plot for the determination of the decay amplitude $\Phi_{B_q}^{ren}$ from a fit to the plateau for light valence quark $q=0.004$ on the $32^3$ ensemble with $a m_\text{sea}^l =0.006$.}
\label{Fig:PhiBq}
\end{minipage}
\hfill
\begin{minipage}{0.48\textwidth}
\vspace{-5mm}
\includegraphics[scale=0.38]{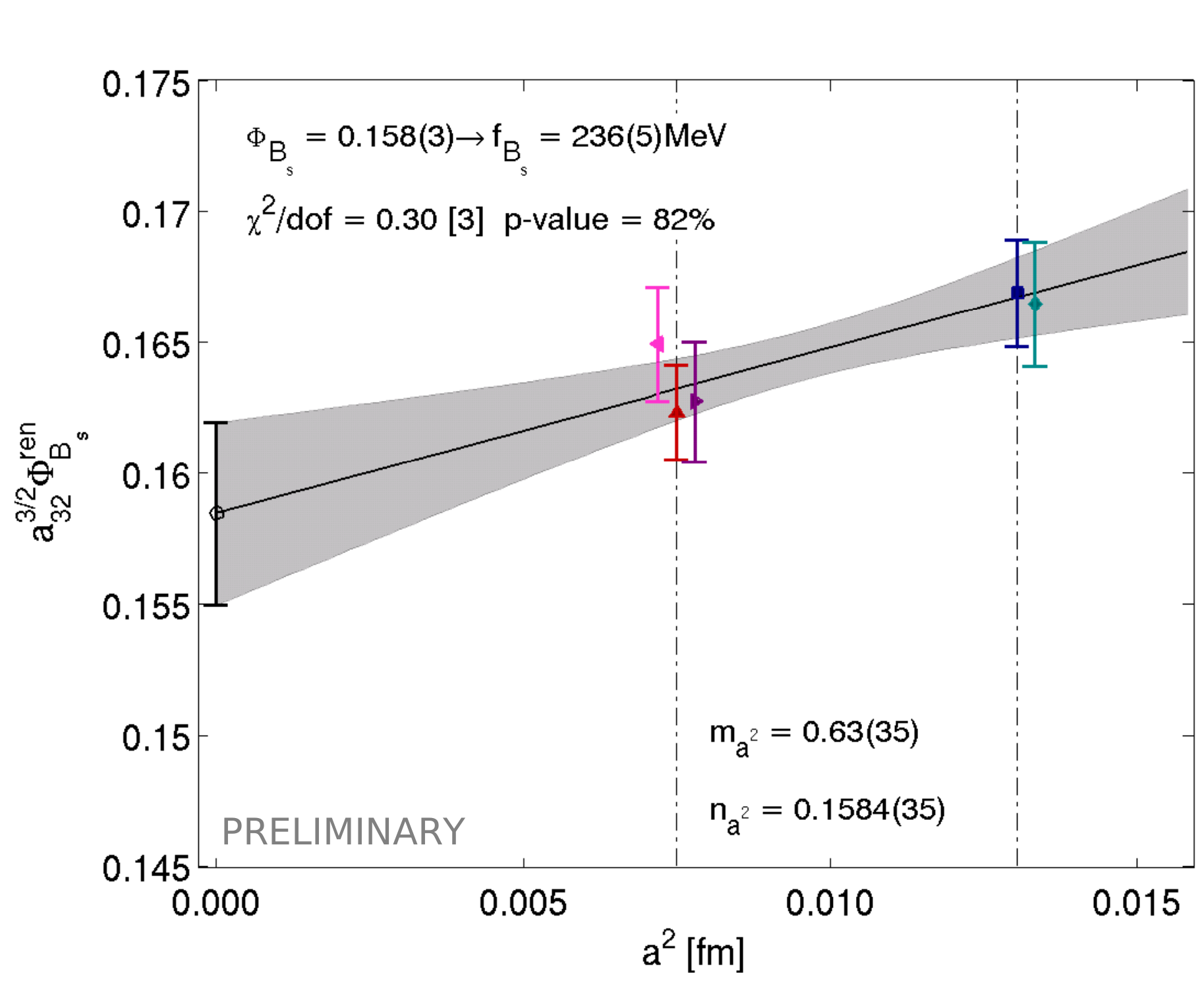}
\caption{Continuum extrapolation of $\Phi_{B_s}$.  The different colored points at each lattice spacing correspond to different sea-quark ensembles, and are horizontally offset for clarity.}
\label{Fig:PhiBs}
\end{minipage}
\end{figure}

The decay constants and the ratio are obtained by first fitting plateaus of the $O(a)$-improved and renormalized decay amplitudes, $\Phi_{B_q}^\text{ren} = Z_\Phi \Phi_{B_q}^\text{imp}$, for all six valence quark masses on our five ensembles. An example for $q=0.004$ on the $32^3$ ensemble with $am_\text{sea}^l = 0.006$ is given in Fig.~\ref{Fig:PhiBq}. We determine $f_{B_s}$ by performing a linear interpolation of the three strange-like data points to the physical value of the strange quark mass. Then we extrapolate the interpolated results on the five ensembles to the continuum with a function that is linear in $a^2$ (motivated by the leading scaling behavior of the light-quark and gluon actions) and independent of sea-quark mass and obtain $f_{B_s} = 236(5)$~MeV (statistical error only).
%\clearpage

\begin{table}[t]
\flushright
\begin{minipage}{0.52\textwidth}
\vspace{-88.5mm}
\caption{Preliminary results for the nonperturbative determination of the flavor-conserving renormalization factor $Z_V^{bb}$ with statistical errors only. Averaging values at the same lattice spacing we compare to the values obtained from a calculation using 1-loop mean-field improved lattice perturbation theory \cite{Lehner:2012bt}. }
\label{Tab:Zvbb}
\begin{tabular}{clcl}\toprule
$a_{24} m_\text{sea}^l$&~~~$Z_V^{bb}$&$a_{32}m_\text{sea}^l$&~~~$Z_V^{bb}$\\
\midrule
0.005 & 10.037(34) & 0.004&5.270(13)\\
0.010 & 10.042(37) & 0.006&5.237(12)\\
      &            & 0.008&5.267(15) \\ \midrule
Avg.$_{24}$&10.093(25)&Avg.$_{32}$&5.2560(76)\\
PT$_{24}^\text{1-loop}$&10.72&PT$_{32}^\text{1-loop}$&5.725\\
\bottomrule
\end{tabular}
\end{minipage}
%\vspace{80mm}
\end{table}

\begin{figure}[t]
%\centering
\begin{minipage}{0.49\textwidth}
\includegraphics[scale=0.38]{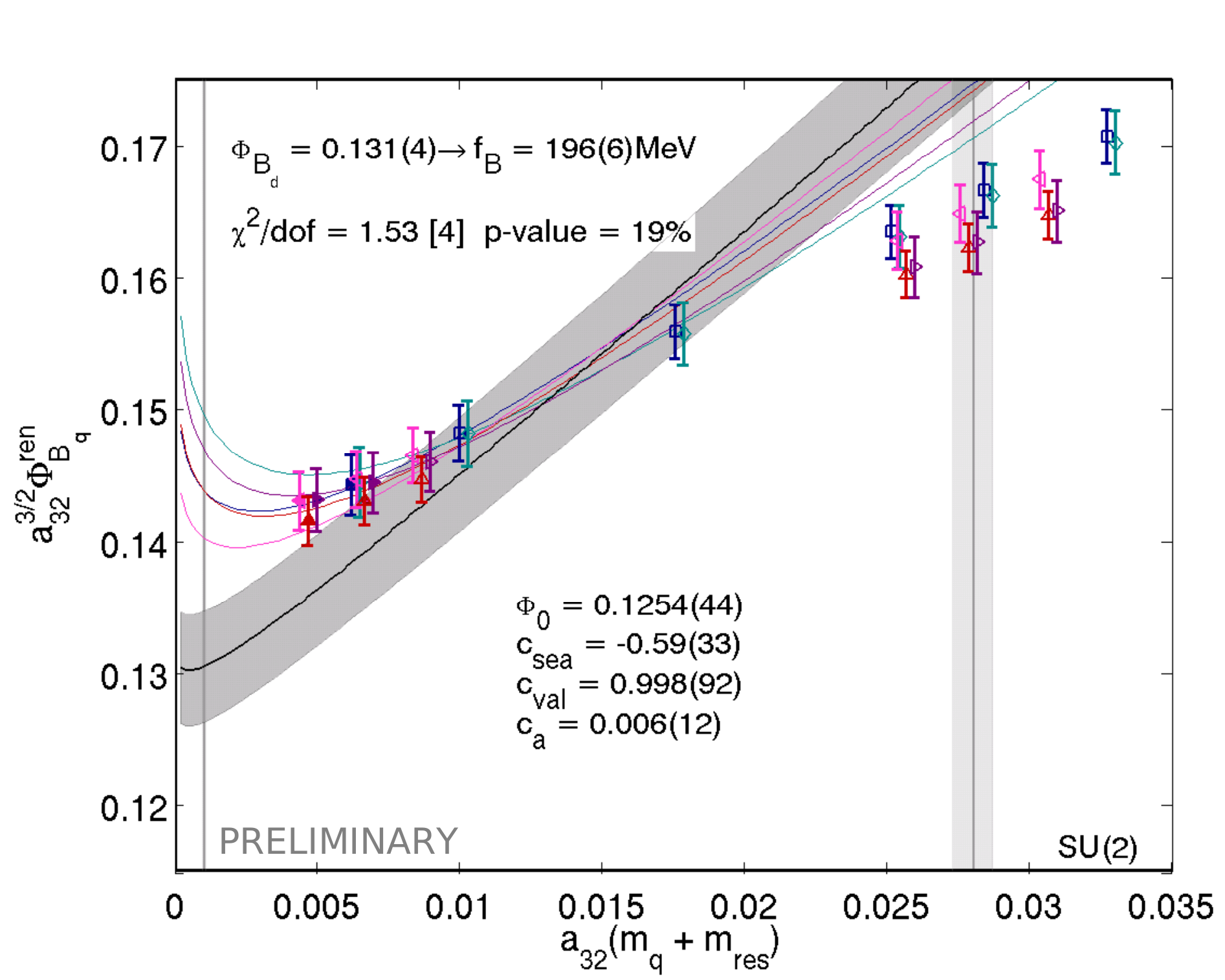}
\end{minipage}
\hfill
\begin{minipage}{0.49\textwidth}
\includegraphics[scale=0.37]{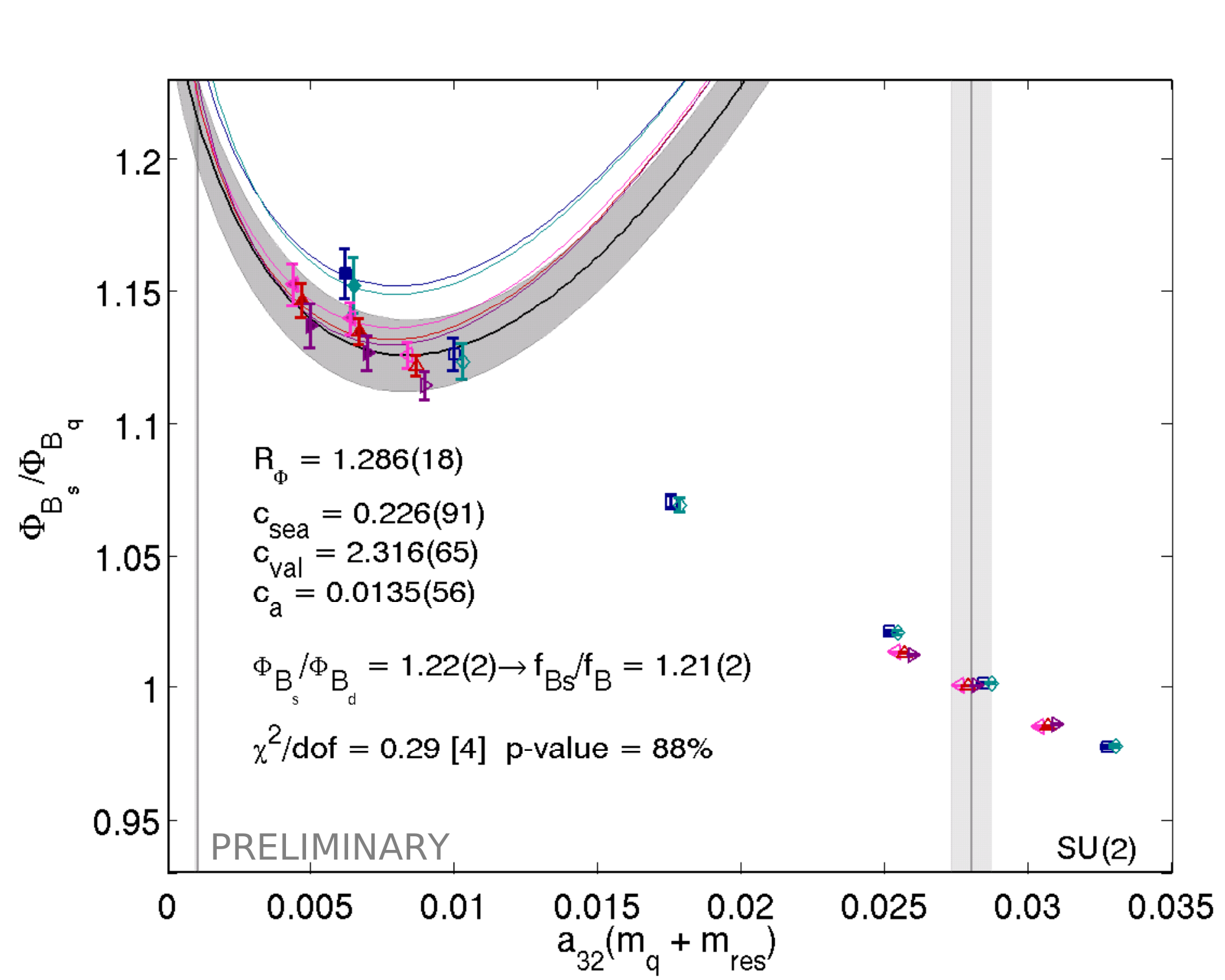}
\end{minipage}
\caption[width=\textwidth]{Chiral-continuum extrapolation of $\Phi_{B_q}$ (left) and $\Phi_{B_s}/\Phi_{B_q}$ (right). For better visibility some data points are plotted with a small horizontal offset. Only the filled data points are included in the fit.  The vertical gray bands indicate the physical values of the $u/d$- and $s$-quark masses \cite{Allton:2008pn,Aoki:2010dy}.}
\label{Fig:PhiB}\vspace{-1mm}
\end{figure}

The physical value of the decay constant $f_B$ and the ratio $f_{B_s}/f_B$ are obtained from a combined chiral-continuum extrapolation using next-to-leading order SU(2) heavy meson chiral perturbation theory (HM$\chi$PT) \cite{Goity:1992tp,Arndt:2004bg,Aubin:2005aq,Albertus:2010nm}
\begin{align}
\Phi_B &= \Phi_0 \left[1-\chi_\text{SU(2)}^{f_B} + c_\text{sea} m_\text{sea}^l 2B/(4 \pi f)^2 + c_\text{val} m_\text{val} 2B/(4 \pi f)^2  + c_a a^2/(a_{32}^2 4 \pi f)^2 \right], \\
\text{and\hspace{9mm}}& \nonumber\\ 
\Phi_{B_s}/\Phi_B &= R_{\Phi} \left[1-\chi_\text{SU(2)}^\text{ratio}  + c_\text{sea} m_\text{sea}^l 2B/(4 \pi f)^2 + c_\text{val} m_\text{val} 2B/(4 \pi f)^2  + c_a a^2/(a_{32}^2 4 \pi f)^2 \right].
\end{align}
The chiral logarithms, $\chi_\text{SU(2)}^{f_B}$ and $\chi_\text{SU(2)}^\text{ratio}$, are nonanalytic functions of the pseudo-Goldstone meson masses, and are given in the appendix of reference \cite{Albertus:2010nm}. Our preliminary results are shown in Fig.~\ref{Fig:PhiB}. The fits are performed including partially-quenched data on all five sea-quark ensembles, but with valence-quark masses restricted to be below $M_\pi^\text{val} < 350$ MeV. These extrapolations give us a preliminary value of $f_B = 196(6)$ MeV and a SU(3) breaking ratio of $f_{B_s}/f_{B_q}$ = 1.21(2). Again only statistical uncertainties are quoted. We are finalizing our budget of systematic errors which will also include, e.g., heavy quark discretization errors. All our preliminary results are in agreement with the literature in particular if taking into account that systematic errors will be added.

\section{Outlook}
We hope to complete and publish our analysis for $f_B$, $f_{B_s}$ and their ratio $f_{B_s}/f_B$ soon. We anticipate that our largest source of error will be from the chiral-continuum extrapolation. In the future, we will take advantage of the new M\"obius domain-wall ensembles generated by the RBC-UKQCD collaboration which feature simulations at the physical pion mass.\vspace{3mm}

\section*{Acknowledgments}
We thank our colleagues of the RBC and UKQCD collaborations for useful help and discussions. Numerical computations for this work utilized USQCD resources at Fermilab, in part funded by the Office of Science of the U.S.~Department of Energy, as well as computers at Brookhaven National Laboratory and Columbia University. O.W. acknowledges  support at Boston University by the U.S. DOE grant DE-SC0008814.

%\begin{thebibliography}{99}
%  \bibitem{...} ....
%\end{thebibliography}
%\clearpage
{\small
\bibliography{B_meson}
\bibliographystyle{apsrev4-1}
}
\end{document}